\documentclass[reprint,rsi,aip]{revtex4-1}

\usepackage{graphicx}
\usepackage[colorlinks=true, citecolor=blue]{hyperref}
\usepackage{color}
\usepackage{textcomp}
\usepackage{amsmath}

\begin{document}

\graphicspath{{./Figures/}}

\title{Global Feed-Forward Vibration Isolation in a km scale Interferometer}

\author{Ryan DeRosa}
\affiliation{Department of Physics and Astronomy, Louisiana State University, Baton Rouge, Louisiana, 70803}
\author{Jennifer C Driggers}
\affiliation{LIGO Laboratory, California Institute of Technology, Pasadena, California, 91125}
\author{Dani Atkinson}
\affiliation{LIGO Hanford Observatory, 127124 N Route 10, Richland, Washington, 99354}
\author{Haixing Miao}
\affiliation{LIGO Laboratory, California Institute of Technology, Pasadena, California, 91125}
\author{Valery Frolov}
\affiliation{LIGO Livingston Observatory, 19100 LIGO Lane, Livingston, Louisiana 70754}
\author{Michael Landry}
\affiliation{LIGO Hanford Observatory, 127124 N Route 10, Richland, Washington, 99354}
\author{Joseph Giaime}
\affiliation{Department of Physics and Astronomy, Louisiana State University, Baton Rouge , Louisiana, 70803}
\affiliation{LIGO Livingston Observatory, 19100 LIGO Lane, Livingston, Louisiana 70754}
\author{Rana Adhikari}
\affiliation{LIGO Laboratory, California Institute of Technology, Pasadena, California, 91125}


\begin{abstract}
Using a network of seismometers and sets of optimal filters, we implemented a feed-forward control technique 
to minimize the seismic contribution to multiple interferometric
degrees of freedom of the LIGO interferometers. 
The filters are constructed by using the Levinson-Durbin recursion relation to approximate the optimal Wiener filter. 
By reducing the RMS of the interferometer feedback signals below $\sim$10~Hz, we 
have improved the stability and duty cycle of the joint network of gravitational wave detectors. 
By suppressing the large control forces and mirror motions, we have dramatically reduced the rate of non-Gaussian transients in the gravitational wave signal stream.
\end{abstract}

\maketitle
\section{Introduction}
\label{sec:intro}
\begin{figure}[h]
	\centering
	\includegraphics[width=\columnwidth, trim=0 0 -27mm 0, clip]{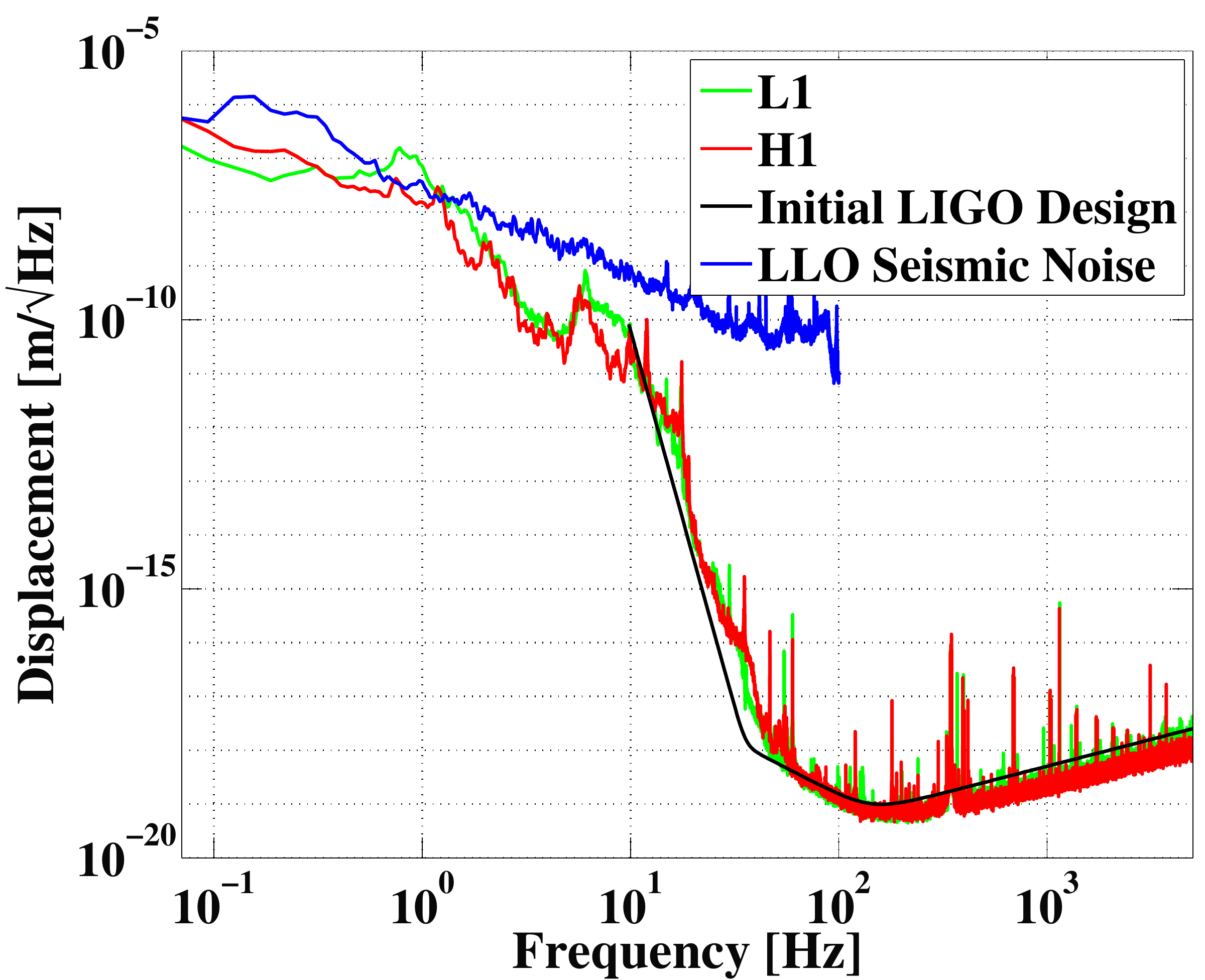}
	\caption{Typical displacement sensitivity for the LIGO interferometers. L1 and H1
	refer to the detectors at Livingston and Hanford, respectively. The black curve is the initial LIGO
	design sensitivity, which was not specified below 10 Hz.}
\label{fig:darm_curve}
\end{figure}

The Laser Interferometer Gravitational-wave Observatory (LIGO)~\cite{PF:RPP2009} consists of detectors located in 
Livingston, Louisiana (LLO) and Hanford, Washington (LHO).
The goal of the experiment is to measure minute disturbances in the relative positions of test mass 
mirrors generated by gravitational waves, using laser interferometry. 
Passing gravitational waves are expected to produce a quadrupolar strain in the local space-time
metric, resulting in a differential phase shift between the laser fields traveling in the two arms of the interferometer.
Potential astrophysical sources of gravitational waves include supernovae and compact binary inspirals.
The expected strain from such sources requires a detector sensitive to displacements on the
order of $10^{-19}$ m/$\sqrt{\textnormal{Hz}}$ over a few km baseline, at frequencies 
ranging from $\sim$50~Hz to 7~kHz.
Figure~\ref{fig:darm_curve} shows a typical displacement sensitivity for the LIGO interferometers. 

\begin{figure}[h]
	\centering
	\includegraphics[width=\columnwidth]{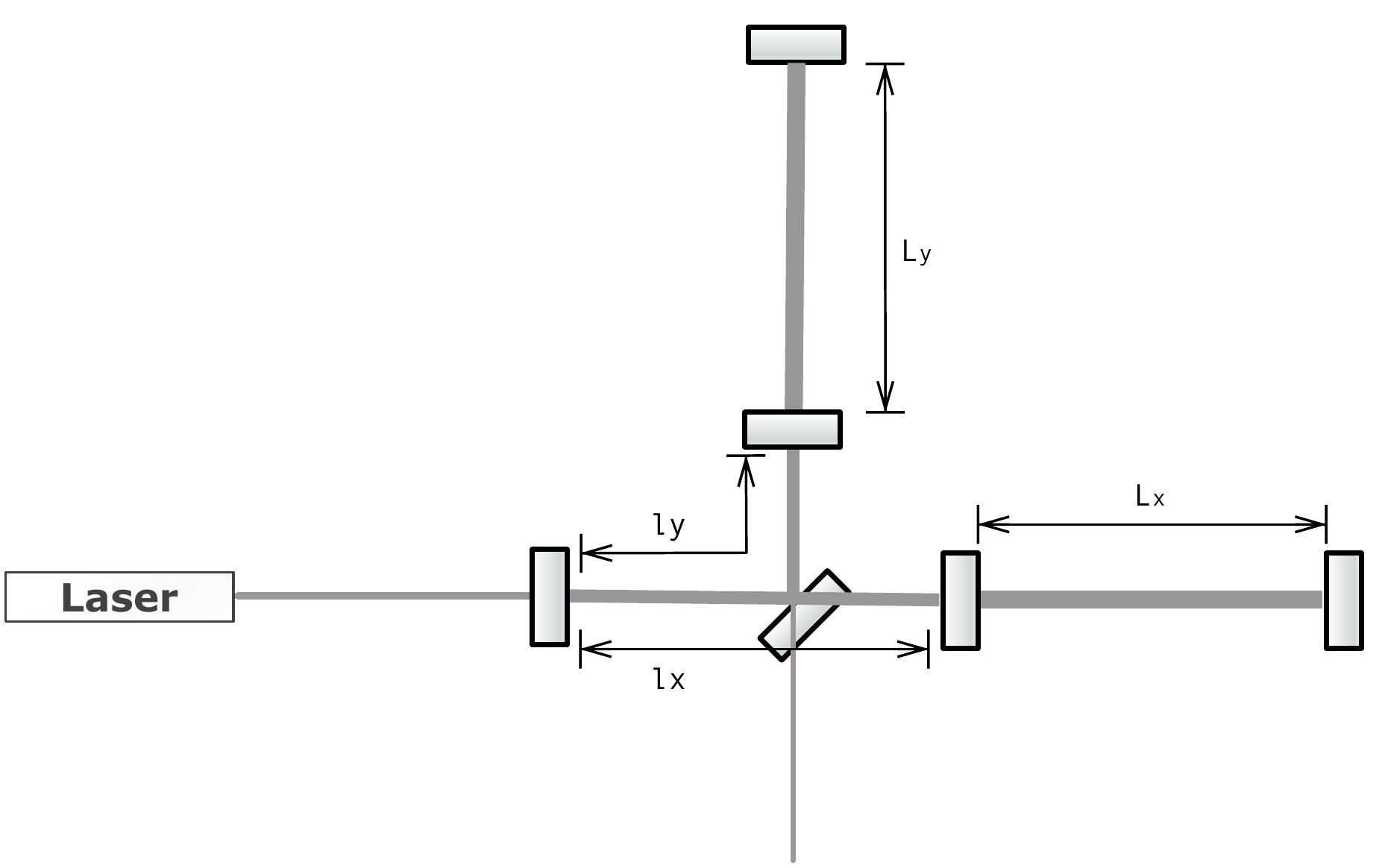}
	\caption{Schematic layout of a LIGO interferometer. The Fabry-Perot arm cavities have
	lengths $L_x \simeq L_y \simeq 4 km$.}
     \label{fig:IFO}
\end{figure}

Using Fabry-Perot cavities for the interferometer arms as well as power recycling~\cite{meers:PR}
helps improve the detector's sensitivity.
Figure~\ref{fig:IFO} shows a simplified schematic of a LIGO interferometer.
The motion of the ground is many orders of magnitude above the required sensitivity, so the mirrors
are seismically isolated, using both passive and active techniques.
The cavities are held on optical resonance by suppressing external disturbances with a family of length and alignment feedback control loops.
Suspending the mirrors as pendulums provides passive vibration isolation above the pendular resonance,
which is arranged to be $\sim$1~Hz.
The support point of the pendulum is attached to a passive in-vacuum seismic isolation stack, consisting 
of four mass and spring layers.
This payload is attached to tubes which exit the vacuum and are supported by
an external active seismic isolation system. 
Small magnets are glued to the backs of the mirrors to allow for actuation via magnetic fields
generated by currents flowing in nearby wire coils.
Above $\sim$50~Hz, where ground vibrations have been sufficiently suppressed, the detector
is fundamentally limited by thermal noise in the mirrors and seismic isolation systems and by
photon shot noise.

There are four length degrees of freedom that need to be
controlled. They are defined in terms of differential and common length changes,

\begin{equation*} L_+ = \frac{L_x + L_y}{2} \end{equation*} 
\begin{equation*} L_- =  L_x - L_y \end{equation*} 
\begin{equation*} l_+ = \frac{l_x + l_y}{2} \end{equation*} 
\begin{equation*} l_-  = l_x - l_y \end{equation*} 

where $L_x$, $L_y$, $l_x$, and $l_y$ are as shown in Figure~\ref{fig:IFO}.
An RF modulation scheme~\cite{drever:PDH} is used to generate control signals by relating position and 
angle fluctuations of the test mass mirrors to power fluctuations measured by photodetectors at various locations around the interferometer~\cite{fritschel:ISC}.
The detector is brought to and held at its operating point by sequentially bringing cavities into resonance using dynamically calculated error signals~\cite{mevans:locking}, a process called lock acquisition. 

Precision measurement of the differential arm length degree of freedom,  called $L_-$ or DARM, 
enables the detector to find potential gravitational waves. 
The common (mean) arm length, $L_+$ or CARM, is also sensitive to gravitational waves but is 
used as a reference to stabilize the frequency of the laser.
When the arm lengths are set such that the antisymmetric port is near the dark fringe most of the 
input power is sent back towards the laser, such that the interferometer acts as a highly reflecting mirror. 
By controlling the common power recycling cavity length, $l_+$ or PRC, another resonant 
cavity can be formed between the power recycling mirror and the Fabry-Perot Michelson, increasing 
the power on the beamsplitter and in the arms.
In order to maintain the Michelson fringe offset required for this configuration the short Michelson
cavity length, $l_-$ or MICH, must also be controlled. 

Of the external disturbances that the control systems must overcome, one of the most
problematic is the persistent and large 0.1 - 0.3 Hz double-frequency
'microseismic' peak, generated by wave activity in the oceans~\cite{daw:MSFF}. 
At the microseismic peak the amplitude is on the order of several 
\textmu m/$\sqrt{\textnormal{Hz}}$, but the level of seismicity present at the observatories
varies widely (by a factor of $\sim$10)~\cite{daw:SE}.
The relative displacement of the test masses in this band is of the same order as the horizontal displacements
caused locally by the microseism.
These levels of seismic noise present a host of problems, including:

\begin{itemize}

\item Some amount of the translational ground motion is converted to angular motion of the mirrors due to
cross-couplings in the mechanical dynamics of the seismic isolation systems.
Imperfect balancing of the mirror actuators also converts some of the length control signals into angular motion.
These misalignments create fluctuations in the power circulating throughout the cavities, driving feedback loops
towards instability.
Several of the angular control loops have bandwidths of only a fraction of a Hz.
Even in the event of single arm locks the full interferometer may not hold resonance due to angular motion
in the power recycling cavity.

\item Large control forces at low frequency can generate excess noise in the
signal band through non-linearities in the mirror actuation. 
Examples of such non-linearities are saturation of the actuator electronics
and Barkhausen noise in the ferromagnets used to actuate the mirrors. 
Non-Gaussian transients from these mechanisms pollute the gravitational wave data stream as background events. 

\item Some amount of light is scattered from each mirror surface, and subsequently re-enters
the readout path after reflecting off of vibrating surfaces.
When the relative motion of the mirror and the other surfaces in the scattering path is larger than the 
laser wavelength ($\lambda \sim 1$ \textmu m), the phase noise introduced by this scattered light is
experienced not only at the frequency of the motion but as broad-band noise up to a cutoff frequency
determined by the relative velocity~\cite{sjw:scatter}.

\end{itemize}

At frequencies where the motion of the ground and the length fluctuations in the cavity are coherent
the seismic noise can be actively subtracted from the cavity motion using a feed-forward signal.
Feed-forward cancellation of the microseism was demonstrated at LLO during the commissioning of the detector,
using piezoelectric actuators and filters created via system-identification~\cite{daw:MSFF}.
Between LIGO's third and fourth science runs (S3 and S4) a new external seismic isolation system, HEPI, 
was installed at LLO, which used feed-forward signals to subtract local ground motion from the 
isolated platform motion with a technique often called `sensor correction' ~\cite{wen:HEPI,hua:FIR}.
Those techniques were used to the absolute motion of each individual platform. 
This paper will describe an improved feed-forward scheme which uses
combinations of seismometers located several km's apart and Wiener filtering to
reduce the differential motion between the platforms, and thereby
quiet interferometric control signals. We show results achieved during LIGO's sixth 
science run (S6) which lasted from July 2009 until October 2010. 
In other work, Wiener filtering has been used for real-time cancellation of phase noise from vibrations in 
cavity stabilized lasers ~\cite{rosenband:FPphasenoise}, and a similar technique was used in a 40 meter long suspended
interferometer ~\cite{jenne:40m}.
\section{Method}
\begin{figure}[h]
	\centering
	\includegraphics[width=\columnwidth]{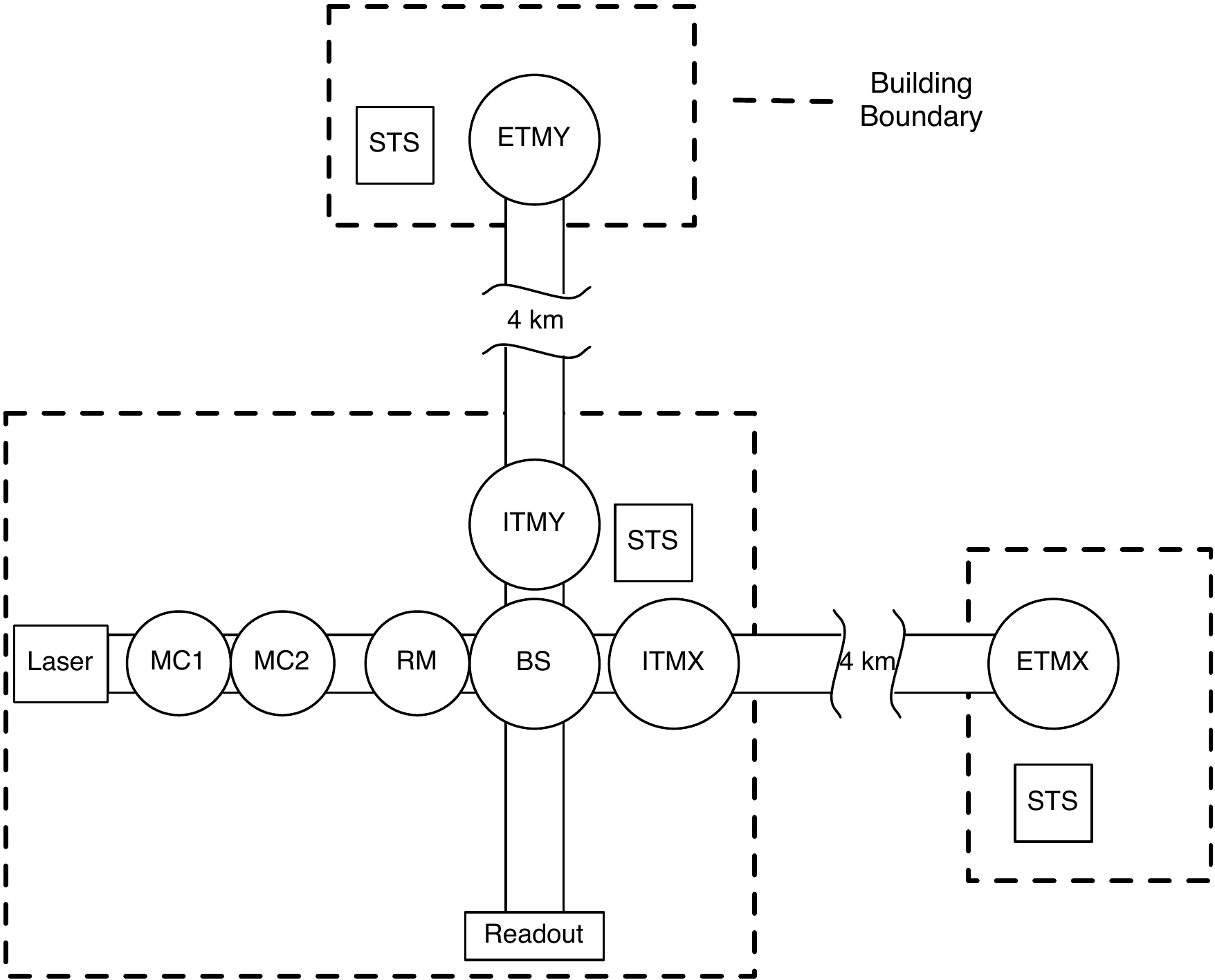}
	\caption{Location of seismometers and vacuum chambers. 
	ETMX/Y and ITMX/Y are the locations of the end and input test masses, respectively. 
	RM is the power recycling mirror, BS the beamsplitter, MC1 and MC2 the mode cleaner chambers, 
	and STS refers to a Streckeisen STS-2 seismometer. At LHO the end stations were equipped with multiple 	single-axis Geotech GS-13 seismometers instead of STS-2's.}
\label{fig:layout}
\end{figure}

In order to optimize the noise subtraction we want to find the Wiener filter~\cite{norbert}, $h$, that processes
the witness seismometer signals, $w$, into the best possible estimate of the targeted
cavity length control signal, $t$. 
We find $h$ by minimizing the mean square error between $t$ and $w$, 

\begin{equation} 
MSE = \frac{1}{2} \sum_i \left[ t_i - \sum_{j=0}^N h_j w_{i-j} \right] ^2 
\end{equation}

\noindent with respect to $h_j$, where $N$ is the filter order (number of taps). 
This is done by setting the derivative,

\begin{equation} \frac{\partial MSE}{\partial h_j} = 0 \end{equation}

\noindent which yields the Wiener-Hopf equations,

\begin{equation} p_i = \sum_{j=0}^N h_j R_{i-j} \end{equation}

\noindent where $R$ is the autocorrelation matrix of the seismometer signal and $p$ is the cross-correlation between the seismometer and the control signals.
The FIR Wiener filters were estimated in MATLAB~\cite{mathworks} using the Levinson-Durbin algorithm~\cite{durbin, huang}, and then fit to a set of IIR coefficients to 
reduce the computational time in the real-time control system.
The conversion from FIR  to IIR was done with the Vectfit~\cite{VectfitPaper} software package.
The intermediate FIR filters were composed of 1000's of taps and processed data with a sample rate of 64~Hz,
allowing for subtraction down to a few 10's of mHz.
Approximately 1 hour of data was used to train each filter.
Each test mass is in the vicinity of at least one seismometer (see Figure~\ref{fig:layout}), which measures 
motion in three perpendicular directions, X, Y, and Z,  where Z is vertical and the two horizontal directions are roughly aligned to the arms of the interferometer. 
The length control is coupled most strongly to motion along the axes of the respective cavities,
but cross-couplings to the other directions are non-negligible.
Regardless of the direction of the witness signal generating a feed-forward correction the actuation is
applied in the direction in which the laser beam is propagating for that particular chamber.

\begin{figure}[h]
   \centering
      \includegraphics[width=\columnwidth]{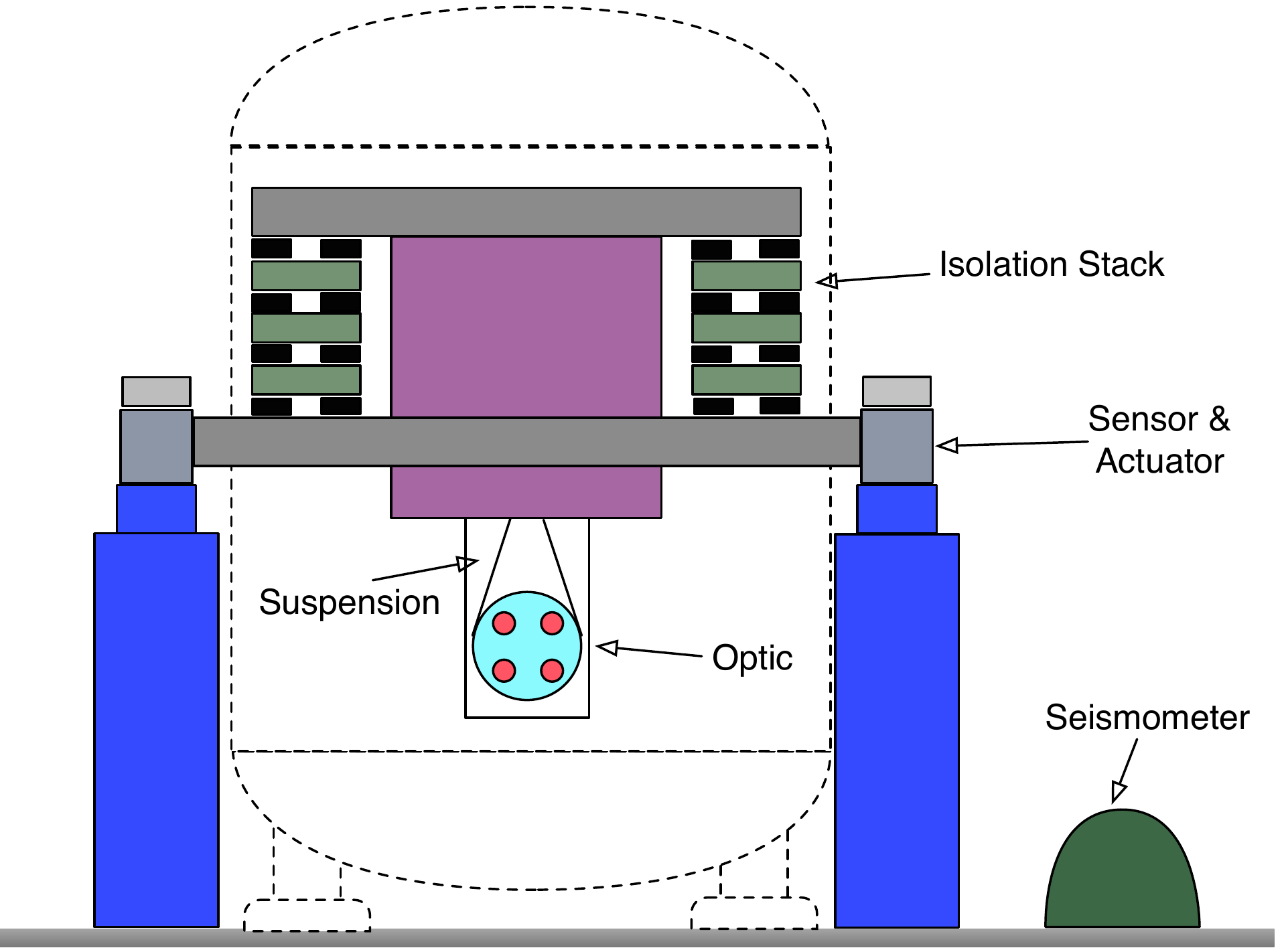}
      \caption{As described in Section~\ref{sec:intro}, the interferometer optic is suspended from the isolation
                 stack. The signals from the ground seismometer are applied
                 just below the stack via piezo (LHO) or hydraulic (LLO)
                 actuators as indicated in the figure. The circles on the optic represent the four
                 electromagnetic actuators on the back of the mirror.
                 The dashed line represents the vacuum chamber.}
   \label{fig:BSCwSUS}
\end{figure}

Offloading control signals to actuators located in the external seismic isolation systems mitigates
several of the problems mentioned in Section~\ref{sec:intro}.
In order to properly subtract the filtered witness signals, the transfer function from our point of actuation
to the cavity control signal must be measured and divided out. 
A diagram showing the relevant pieces of the mechanical structure can be seen in 
Figure~\ref{fig:BSCwSUS}.
The number of mechanical components separating the mirrors from the ground creates
a complicated transfer function with many resonant features.

\begin{figure}[!htb]
   \centering
      \includegraphics[width=\columnwidth]{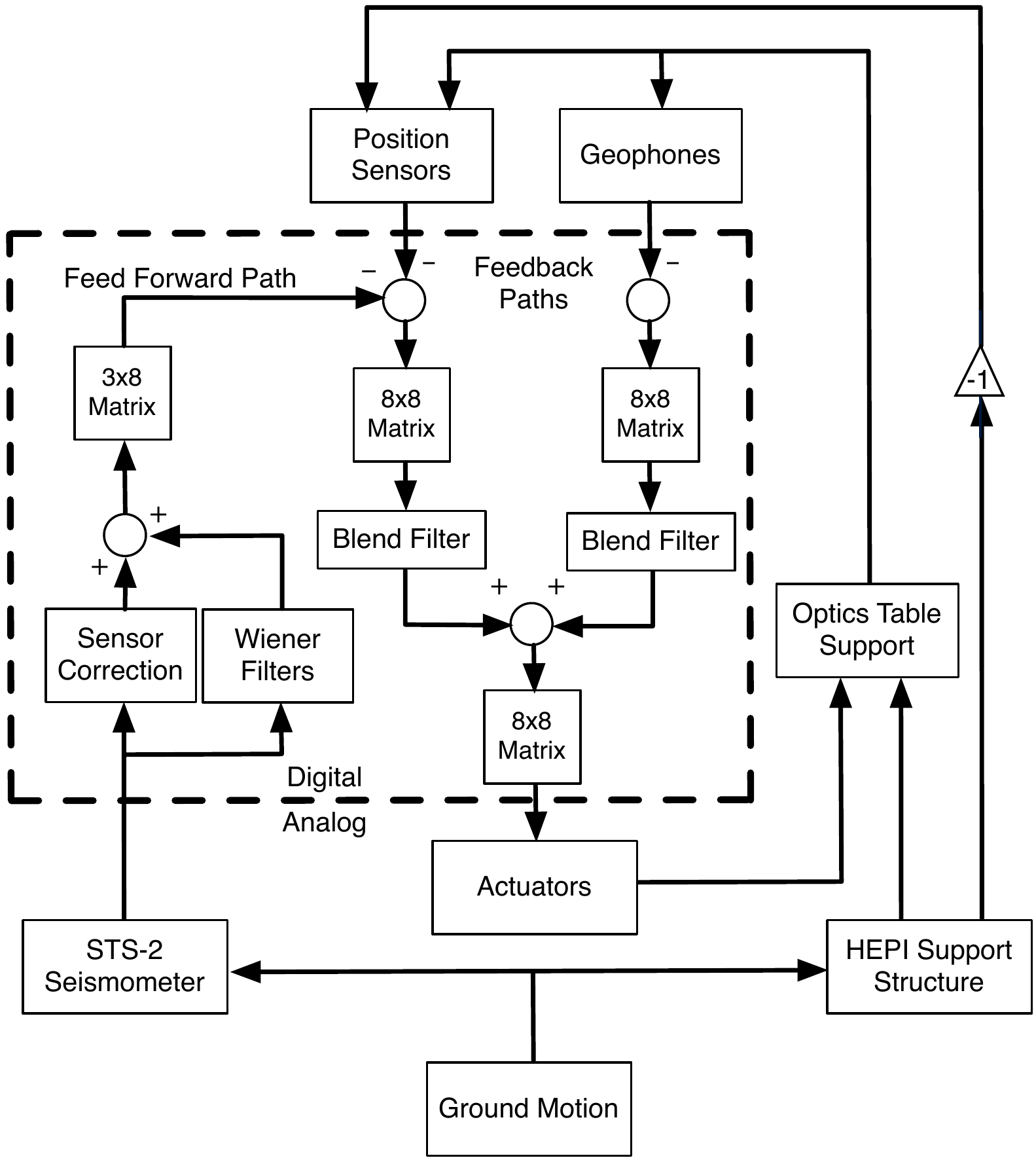}
      \caption{The external isolation systems measure the motions of their payloads using position sensors
      		and geophones, providing feedback via hydraulic or piezoelectric actuators. Ground motion
		is also measured with a nearby seismometer, which is digitally processed and subtracted from
		the position sensor signal, a process called sensor correction. For chambers containing test mass
 		optics this signal is then blended with
		the geophone signal into a so-called `super-sensor' which has good combined sensitivity over
		a wider frequency band than either sensor individually. These `super-sensor' signals are fed to the
		actuators. The presence of conversion matrices indicate a transformation between the different
		coordinate bases used by particular sensors/actuators. The use of four horizontal and four
		vertical position sensors/geophones allows for the sensing and control of 8 degrees of freedom:
		3 translational, 3 rotational, and 2 over-constrained. This schematic is adapted from Figure 3.8 of Wen 		~\cite{wen:HEPI}.}
   \label{fig:hepi_schematic}
\end{figure}

\begin{figure}[htb]
	\centering
	\includegraphics[width=\columnwidth]{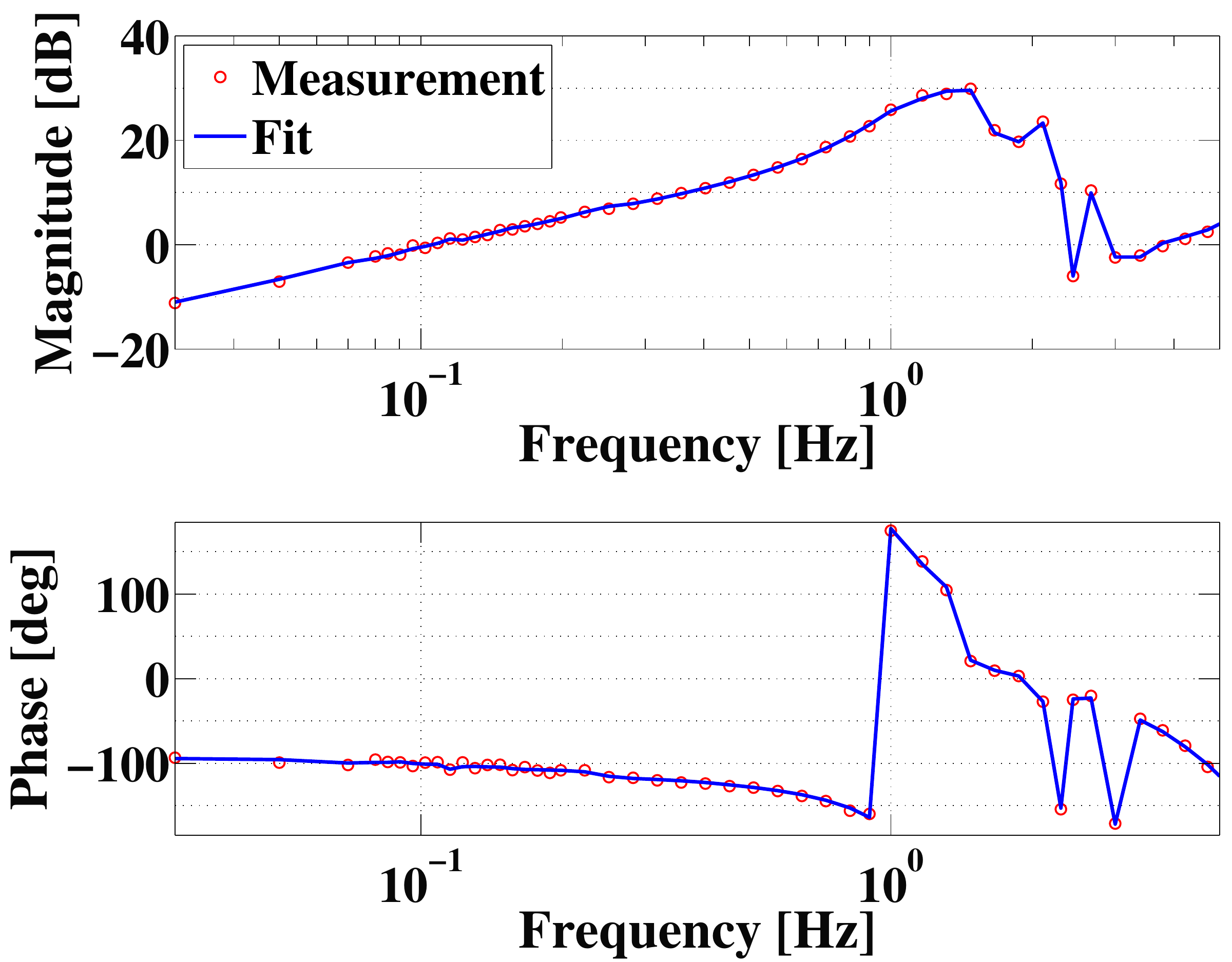}
	\caption{Example plant transfer function, fit using Vectfit. The average statistical error of points
	shown was $\sim$ 2.5\% in magnitude and $\sim 1.5^{\circ}$ in phase.}
\label{fig:example_tf_fit}
\end{figure}

\begin{figure}[htb]
	\centering
	\includegraphics[width=\columnwidth]{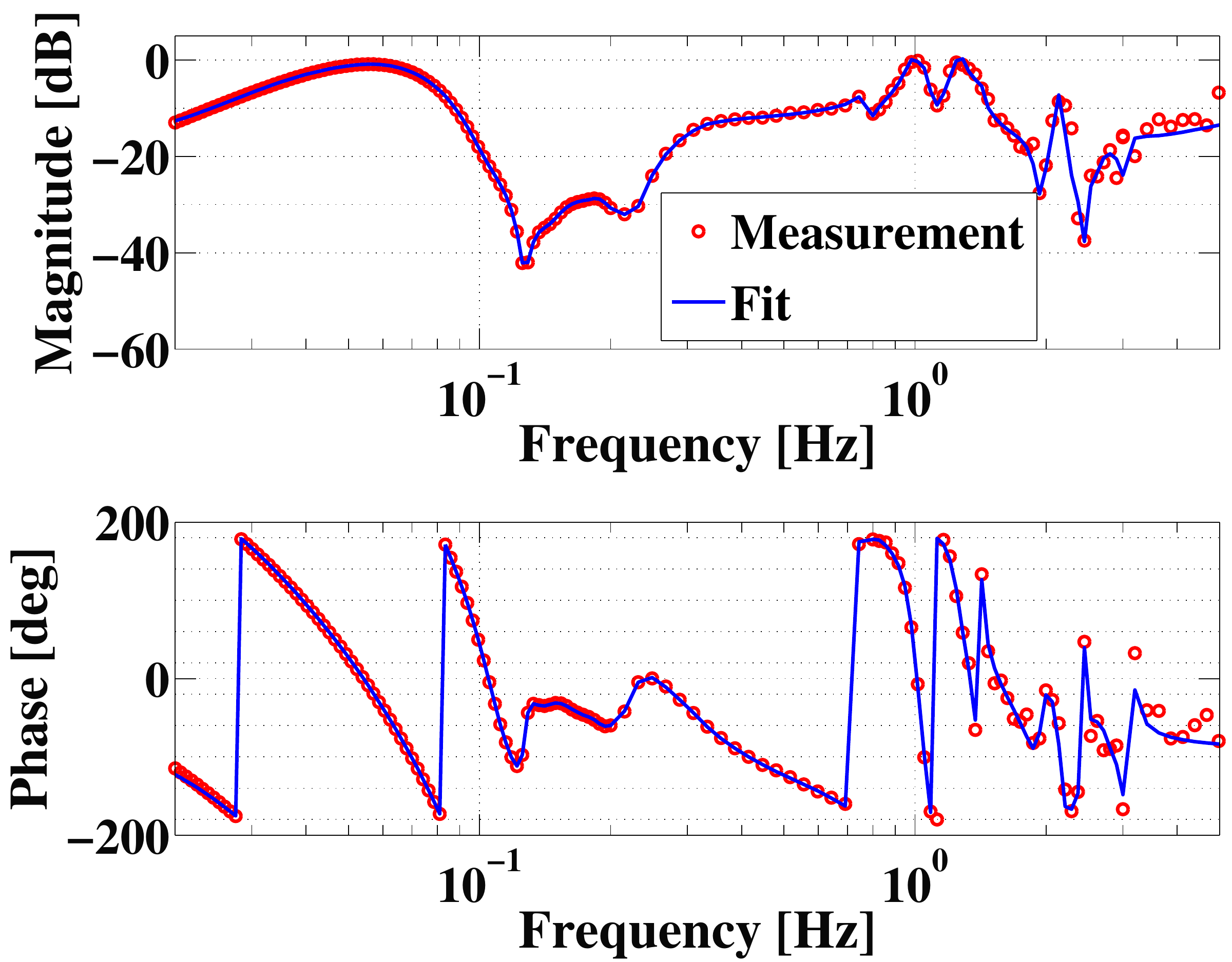}
	\caption{Example Wiener filter, corrected for the mechanical plant response, also fit using Vectfit.}
\label{fig:example_wf_fit}
\end{figure}

At LLO, the active seismic isolation system is HEPI (hydraulic external pre-isolator)~\cite{wen:HEPI,aLIGO:SEI}.
A similar system installed at LHO, PEPI, used piezoelectric actuation.
HEPI is an Advanced LIGO (aLIGO) system which will be in place at both LLO and LHO
in the future. 
Figure~\ref{fig:hepi_schematic} shows a schematic of the HEPI system ~\cite{wen:HEPI}.
The Wiener filter feed-forward path was added in parallel to the existing sensor correction path, which contained hand-tuned filters designed to match HEPI's position sensors to local seismometers on the ground.
Typically this sensor correction reduced the differential motion sensed by the suspended cavities to $\frac{1}{10}$ of the ground motion, in the microseism band.
All results shown are improvements on top of this existing isolation.
The HEPI actuators~\cite{hardham:actuators} provide the ability to move its payload by $\pm$ 700 \textmu m; the maximum range of PEPI was $\pm$ \textmu m.
We excited these actuators while the detector was locked and monitored the mirror motion
to measure the mechanical response. We again used Vectfit to fit these measurements.
An example of these transfer function measurements can be seen in Figure~\ref{fig:example_tf_fit}.
An example Wiener filter can be seen in Figure~\ref{fig:example_wf_fit}. 
Both the Wiener filter and mechanical response are only fit accurately up to few~Hz.
Low-pass filtering was applied to the feed-forward signal to prevent noise injection at higher frequencies.

\section{Results and Limitations}
\begin{figure}[htb]
	\centering
	\includegraphics[width=\columnwidth]{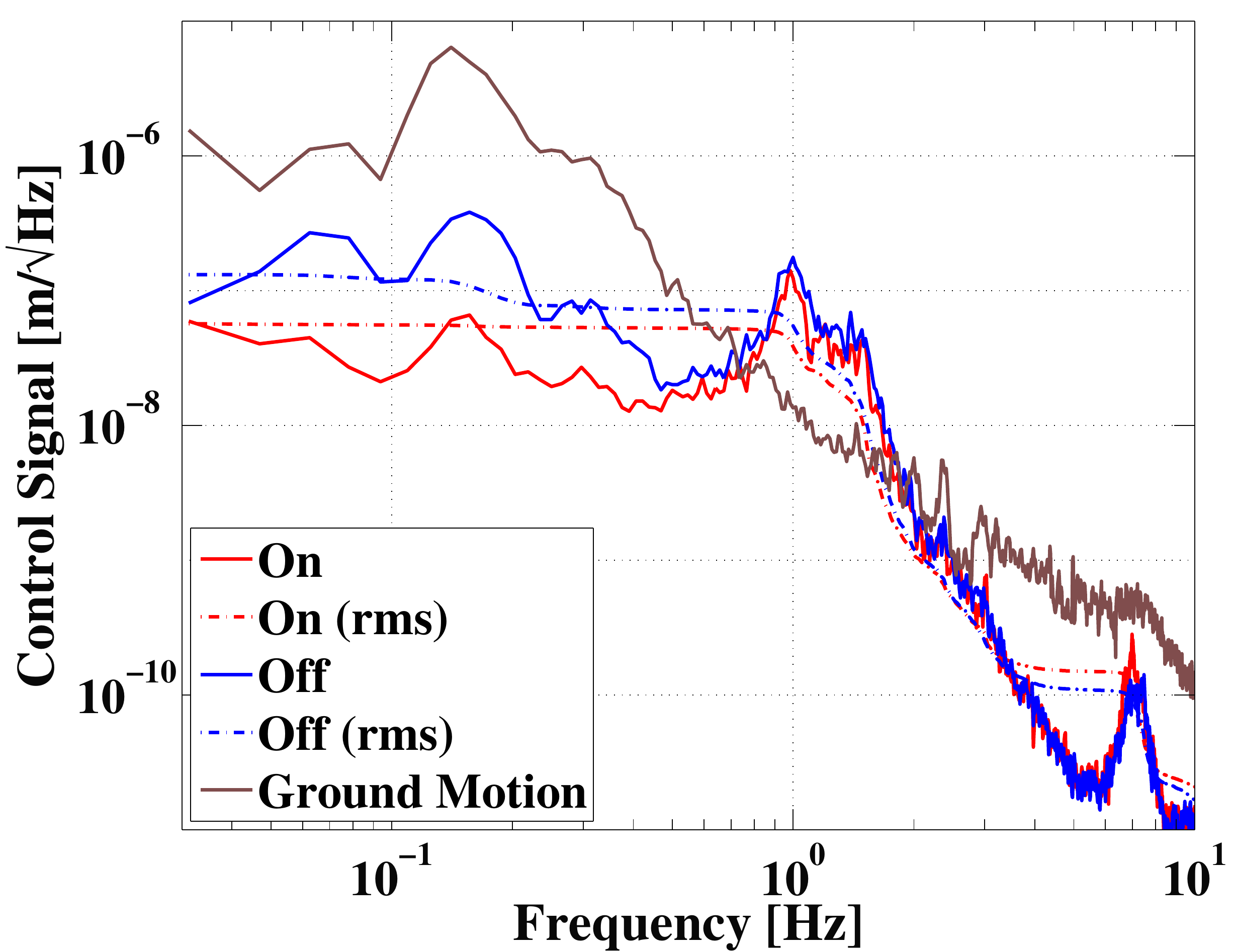}
	\caption{Effect of Wiener filter feed-forward on length control of the mode cleaner at LLO.
	The feed-forward signals were delivered to the actuators along the axis of the beam in MC2's HEPI
	system.}
   \label{fig:llo_mcl}
\end{figure}

Before the laser beam enters interferometer, higher order spatial modes are removed
by a suspended, triangular cavity called the code cleaner (MC). 
The length of the mode cleaner is controlled such that only the TEM00 mode is resonant. 
The cavity is $\sim$12 m long, with its mirrors split between isolated tables in two vacuum tanks, 
MC1 and MC2, as shown in Figure~\ref{fig:layout}. 
At LLO these chambers are equipped with HEPI and served as the initial testbed for the Wiener filter
feed-forward implementation.
Figure~\ref{fig:llo_mcl} shows the control signal holding the mode cleaner on resonance with and without
the Wiener feed-forward signal path enabled. 
A factor of up to $\sim$5 improvement was realized
around the microseismic peak, and the RMS of the signal was reduced by a factor of $\sim$2.
The pre-existing sensor correction filters are left in their nominal state for all results shown in this paper.

\begin{figure}[htb]
	\centering
	\includegraphics[width=\columnwidth]{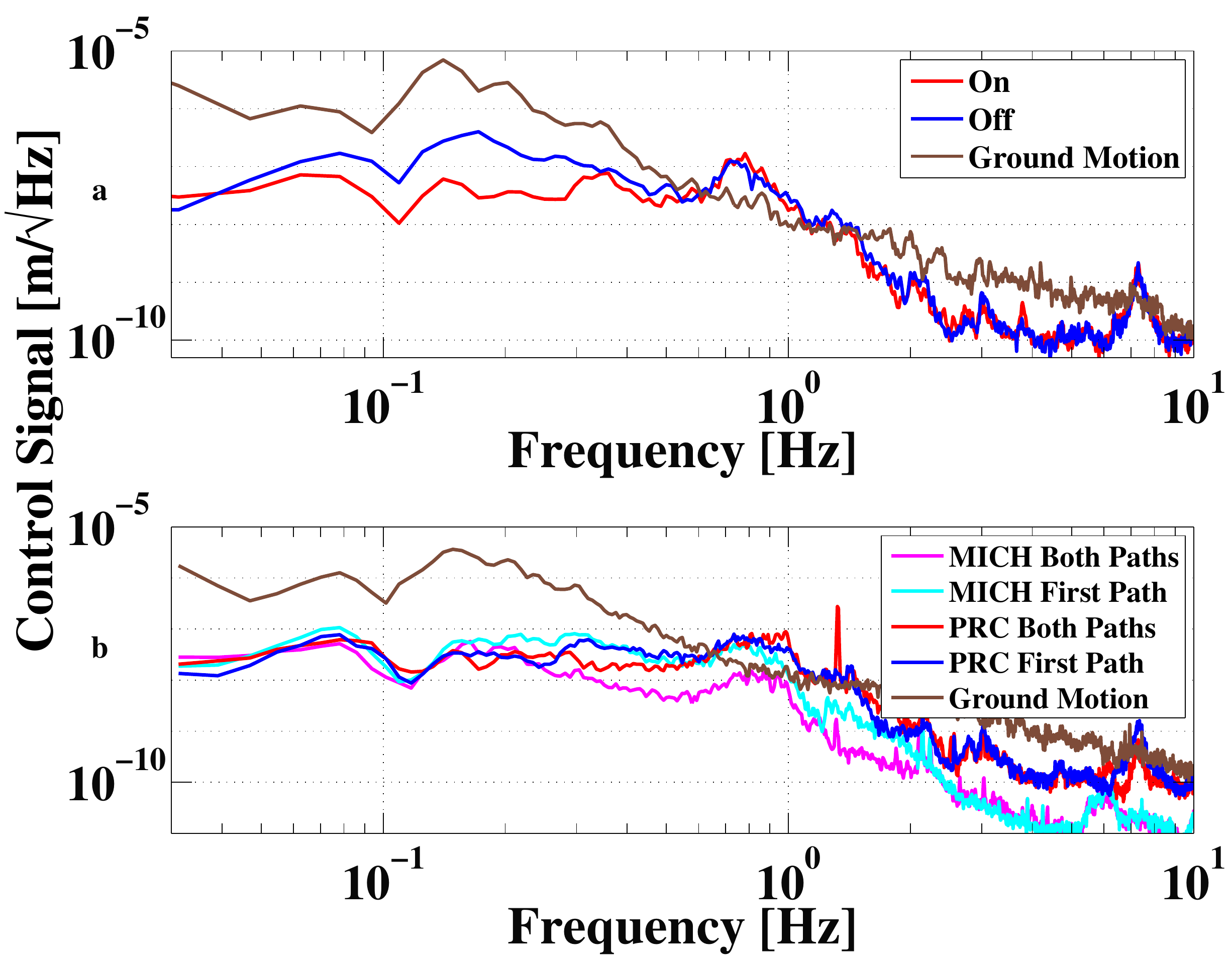}
	\caption{The upper plot shows the reduction in power recycling cavity control signal achieved using Wiener filter feed-forward, improving the isolation of the relative mirror motion to $\simeq \frac{1}{100}$ of the local ground motion. The lower plot shows the implementation of Wiener filter feed-forward on the short Michelson cavity and power recycling cavity simultaneously.}
   \label{fig:llo_prc_mich_ffw}
\end{figure}

Successful implementation of Wiener filter feed-forward on the mode cleaner length control prompted commissioning of feed-forward
paths for the other interferometric control signals, starting with the power recycling cavity control at LLO. 
This work was performed during a time of particularly high microseismic activity which impaired 
the Livingston detector's ability to remain in lock. 
As can be seen in Figure~\ref{fig:llo_prc_mich_ffw}(a), a significant reduction in control signal was achieved,
with the RMS being reduced by more than a factor of 2.
As a result there was also a reduction in the power fluctuations in the recycling cavity, which tended to drive
the interferometer control loops towards instability, and greatly improved the instrument's duty cycle.
This cavity is nearly geometrically degenerate~\cite{andri:PRCoffset} (g-factor $\sim$1) and 
therefore is especially sensitive to misalignments. 

Sending feed-forward signals developed to minimize one cavity control signal may inject noise into
other length controls, since the cavities share mirrors.
To avoid this problem, feed-forward paths were implemented in a serial fashion, such that extra motion
injected into other degrees of freedom could be corrected by the 
Wiener filters constructed for subsequent feed-forward paths.
For example, Figure~\ref{fig:llo_prc_mich_ffw}(b) shows the effect of filters designed to minimize the short
Michelson control signal,
calculated on top of the existing power recycling feed-forward path.
Above $\sim$0.2~Hz the Michelson signal is reduced, with the improved isolation of the power recycling cavity preserved, albeit with some noise injection above $\simeq$1 Hz.
Since the overall RMS is still reduced this noise injection was considered an acceptable tradeoff. 

\begin{figure}[htb]
	\centering
	\includegraphics[width=\columnwidth]{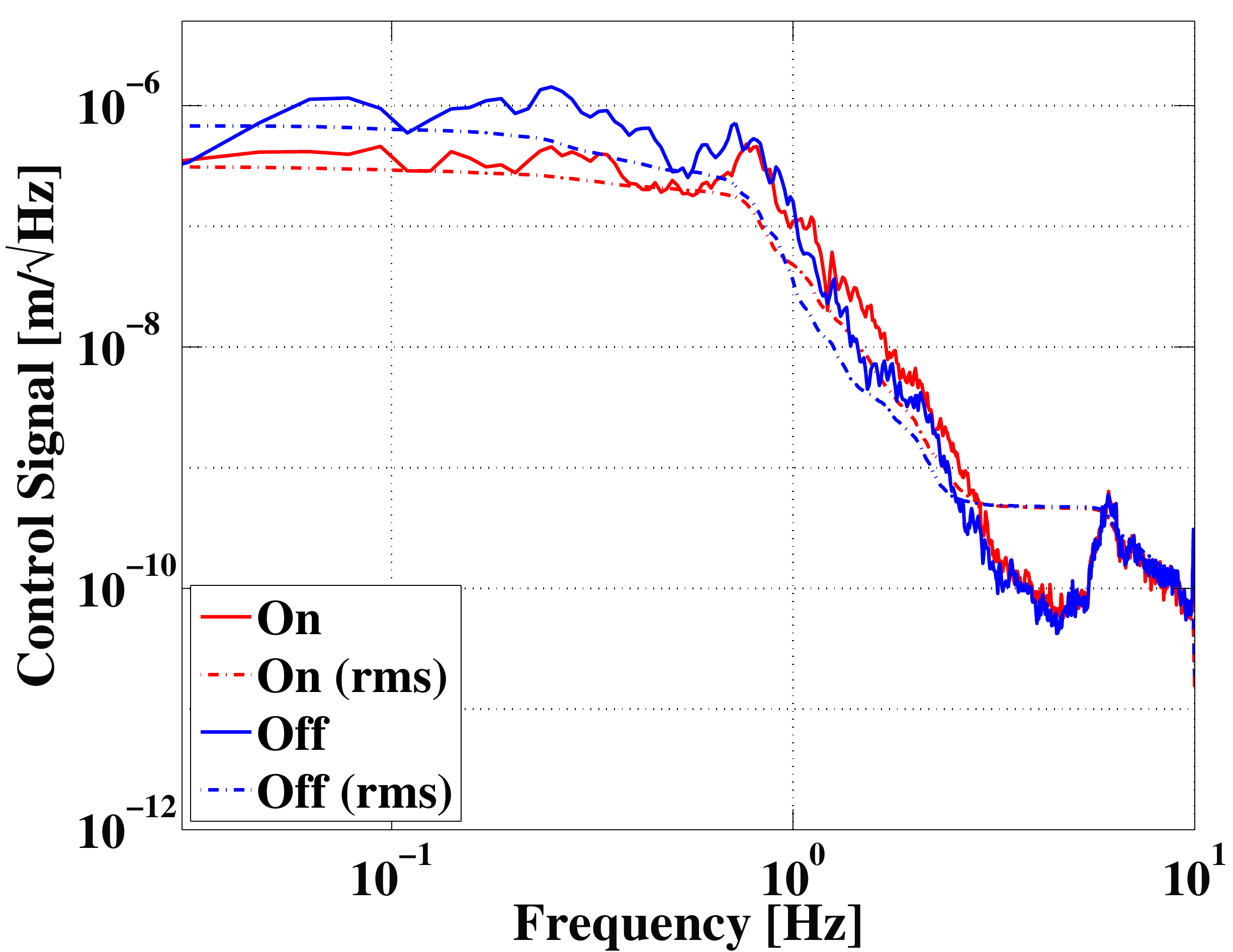}
	\caption{Reduction in differential arm length control from Wiener filter feed-forward signals at LLO.
	The overall RMS of the feedback control was reduced by a factor of almost 3, with some noise injection
	above 1 Hz.}
   \label{fig:llo_darm}
\end{figure}

\begin{figure}[htb]
	\centering
	\includegraphics[width=\columnwidth]{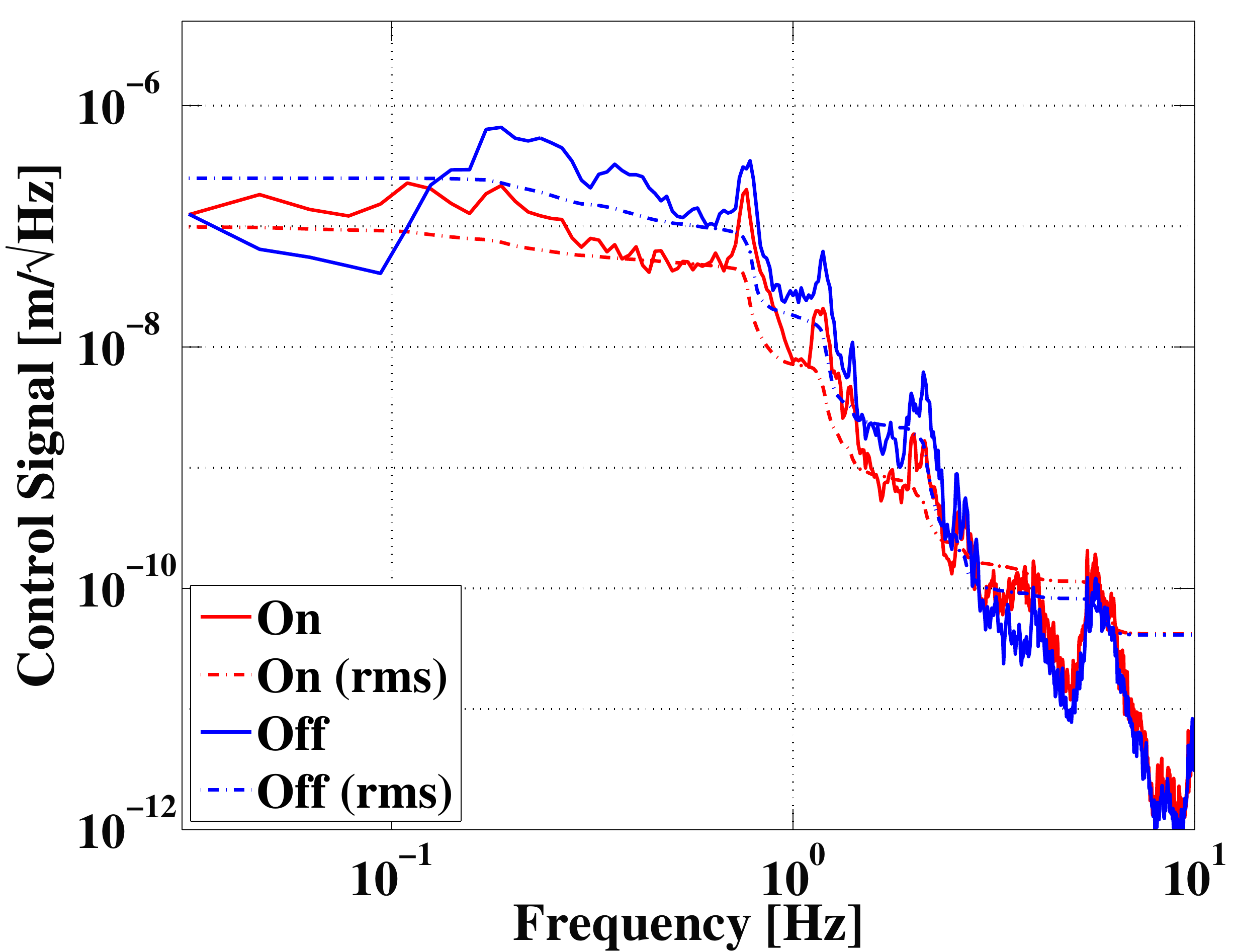}
	\caption{Reduction in differential arm length control from Wiener filter feed-forwad signals at LHO.
	The RMS is reduced in a simliar fashion to LLO, with some excess noise at a few Hz and below 100 mHz.}
   \label{fig:lho_darm}
\end{figure}

Global feed-forward for the differential arm length control was implemented at both observatories with similar
results. 
Since the DARM degree of freedom is sensitive to the motions of mirrors in both the end buildings as well
as the corner station seismometers in all three locations were incorporated into the calculation of the 
optimal feed-forward filter. 
Figures~\ref{fig:llo_darm} and \ref{fig:lho_darm} show that the overall RMS of the control signal is reduced
by a factor of $\sim$2.5 for each interferometer. 
For LLO this improvement in performance was balanced by a slight noise increase above 1 Hz.
Some noise was also injected by Wiener filter feed-forward at LHO, below 0.1~Hz, and above 3~Hz. 

\begin{figure}[htb]
	\centering
	\includegraphics[width=\columnwidth]{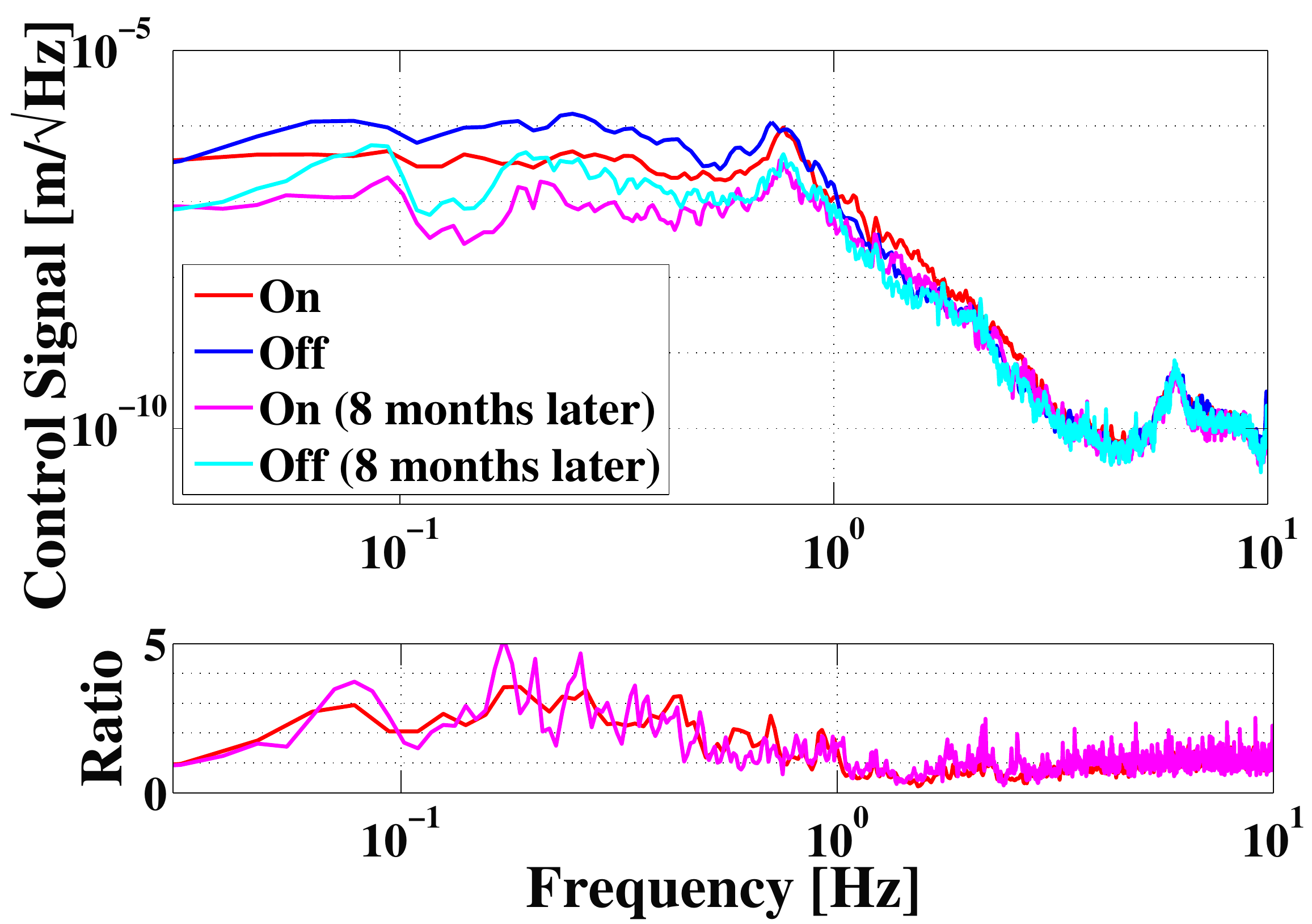}
	\caption{Performance of DARM Wiener filter feed-forward at the time of filter training (upper traces) and 8 		months later (lower traces). The two sets of spectra are from times with significantly different seismic input.
	The lower plot show the ratio of feed-forward off/on for the two time periods.}
   \label{fig:darm_longterm}
\end{figure}

The couplings of ground motion to the feedback signal can potentially be time dependent. 
Since the approach used here is not adaptive, this could lead to a degradation of the filter's
subtraction efficacy over time. 
Figure~\ref{fig:darm_longterm} shows the performance of DARM feed-forward at LLO when first implemented
and 8~months later.
While the overall reduction in RMS motion was originally a factor of $\sim$2.5 the same feed-forward filters
provided $\sim$20\% less isolation after 8~months, reducing the RMS motion by a factor of 2.
There is no obstacle to retraining new Wiener filters to potentially recover subtraction performance, however
if there is a change in the mechanical plant remeasuring the transfer function is a time consuming process, due to the high accuracy required. 

\begin{figure}[htb]
   \centering
      \includegraphics[width=\columnwidth]{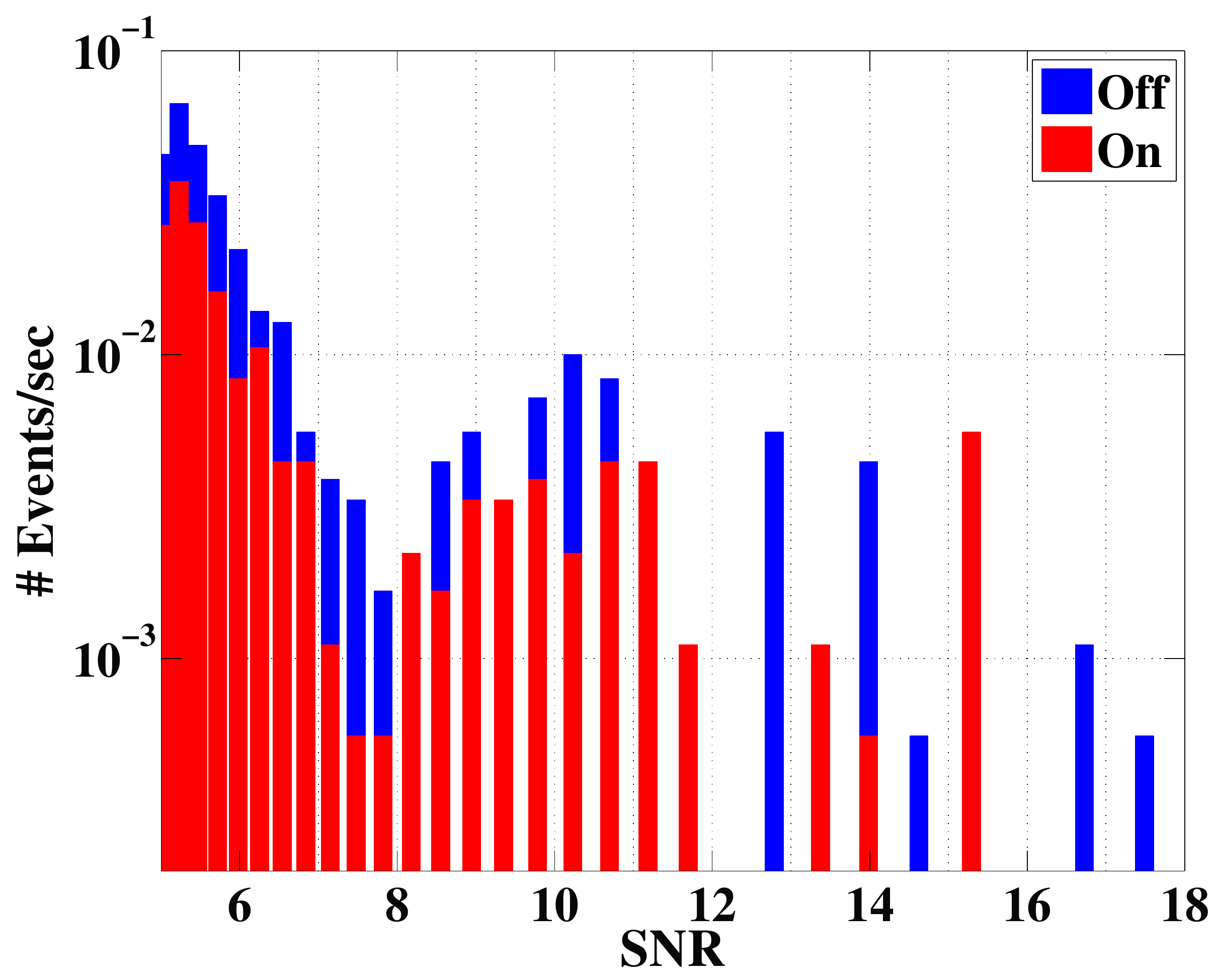}
      \caption{Reduced low frequency control signal in the DARM degree of freedom also results
      		   in reduced transient rates at higher frequencies. This histogram shows the number of
		   transient events per second between 40 and 150 Hz in the L1 interferometer, over 30 minute 
		   stretches with and without global feed-forward.}
   \label{fig:glitchrate}
\end{figure}

As mentioned in Section~\ref{sec:intro}, we have observed that the rate of non-Gaussian events in the gravitational wave band increases when large forces are required to maintain detector lock. 
To measure the effect of improved isolation on the rate of background events, two 30 minute segments of data
were collected for L1, with the DARM Wiener feed-forward first enabled and then disabled.
A templated search, using a sine-Gaussian waveform basis~\cite{chatterji:Q}, was performed and the 
number of triggers per second reported are shown in Figure~\ref{fig:glitchrate}. 
Only triggers with frequencies between 40 and 150 Hz are included, and the threshold SNR was set to 5.
While both distributions show an excess of non-Gaussian events, reducing the low frequency control
signal clearly results in a lower rate of signal band transients. 
The stationary noise floor of the detector did not vary significantly during the time this data was taken.

\begin{figure}[htb]
	\centering
	\includegraphics[width=\columnwidth]{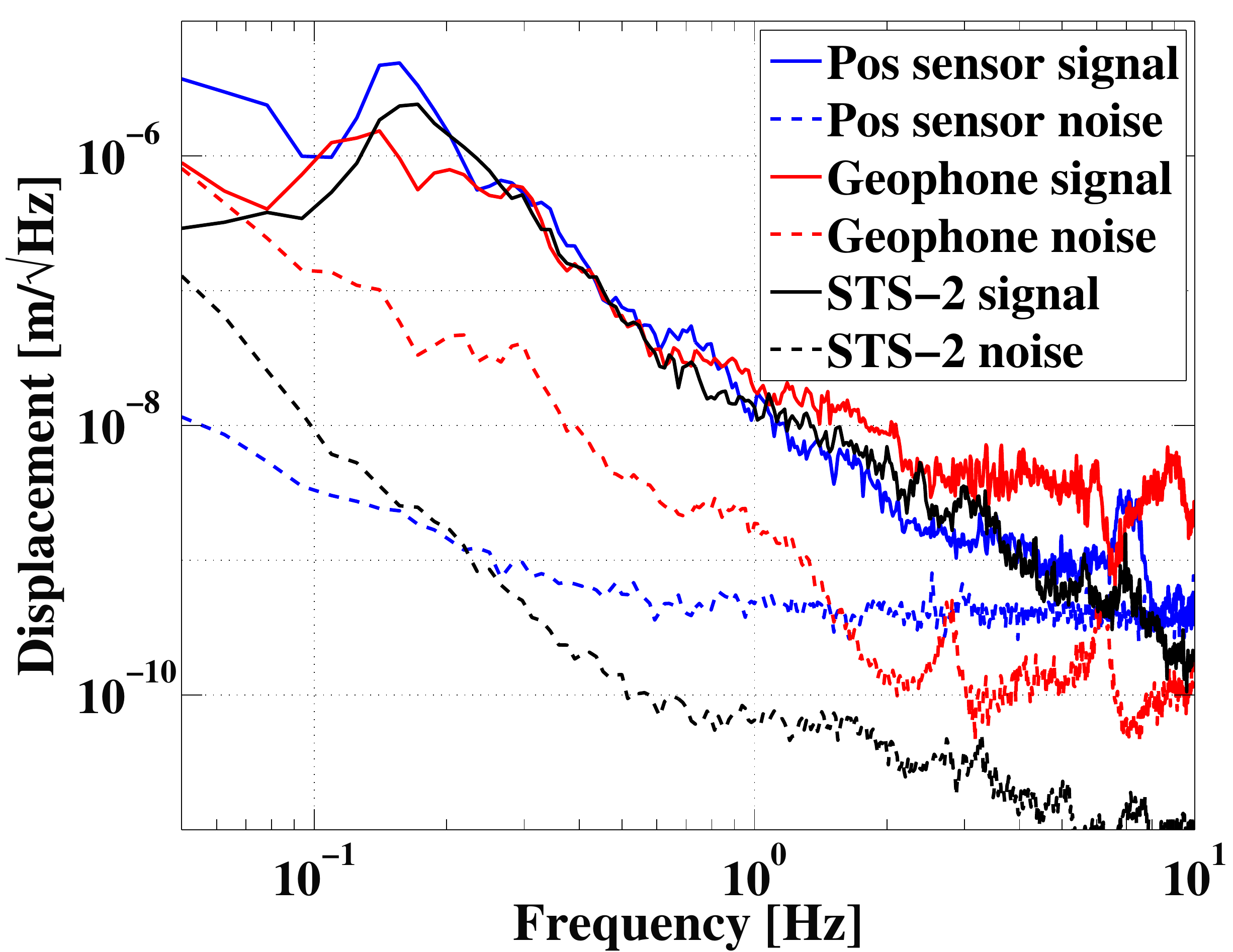}
	\caption{Signal and noise levels for sensors used in the HEPI system. 
	Noise floors shown for the position sensors (Kaman DIT-5200 IPS) and geophones (Mark Products L-4C) 	are the spectra of the horizontal over-constrained modes, i.e., the deformation of the beams connecting the 	HEPI piers. The seismometer (Streckeisen STS-2) noise floor was measured by huddle testing two  		sensors, i.e. positioning them adjacent to each other ($<1$m apart) such that they measure the same 		ground motion and comparing their outputs. These curves only represent the noise floor we are able to 		measure in our installed setup and not necessarily the noise floor of the sensors themselves.}
\label{fig:sensor_noise}
\end{figure}

Fundamentally the ability to provide feed-forward subtraction is limited by the coherence between the witness
signals and the cavity control signal. 
A drawback particular to the scheme detailed in this paper is the lack of witnesses measuring all signals which
contribute to the cavity motion.
Since the seismometers cannot distinguish between translation and tilt, additional sensors such as tilt meters could 
contribute additional isolation through more feed-forward paths~\cite{blantz:rotations}.
There is the potential to be limited by the noise floors of motion sensors,
however at the microseismic peak the signal to noise ratio is very large; see Figure~\ref{fig:sensor_noise}.
Numerical errors can arise in the calculation of the FIR Wiener filter as well as in the fitting of both the IIR
coefficients and the mechanical transfer function.
These inaccuracies also have the potential to limit the feed-forward subtraction. 

\section{Conclusions}
We have demonstrated the ability to use the LIGO detector's active seismic isolation 
system to reduce the relative motion between optics on a 4 km scale through seismometer feed-forward control.
By sending the filtered seismometer signals to the HEPI/PEPI actuators we were able to quiet the relative
motion of the mirrors, offloading control signal from the test mass actuators to the seismic isolation
system.
During times of elevated microseismic activity this allowed for unprecedented interferometer stability. 
The rate of audio band transients is significantly reduced by having less motion and less applied force.
This method will be directly applicable to the second generation gravitational-wave detectors.

\section{Acknowledgements}
We are supported by NSF grant PHY-0905184.
J. C. D. also acknowledges the support of an NSF Graduate Research Fellowship. 
LIGO was constructed by the California Institute of Technology and 
Massachusetts Institute of Technology with funding from the National 
Science Foundation and operates under cooperative agreement PHY- 
0757058.
This article has LIGO Document Number P1000088.

\bibliography{WienerBIB}

\begin{thebibliography}{10}

\bibitem{PF:RPP2009}
BP~Abbott, R.~Abbott, R.~Adhikari, P.~Ajith, B.~Allen, G.~Allen, RS~Amin,
  SB~Anderson, WG~Anderson, MA~Arain, et~al.
\newblock Ligo: the laser interferometer gravitational-wave observatory.
\newblock {\em Reports on Progress in Physics}, 72:076901, 2009.

\bibitem{meers:PR}
B.~J. Meers.
\newblock {\em Phys. Rev. D}, 38:2317--26, 1988.

\bibitem{drever:PDH}
R.W.P. Drever, J.~L. Hall, F.~V. Kowalski, J.~Hough, G.~M. Ford, A.~J. Munley,
  and H.~Ward.
\newblock {\em Appl. Phys. B}, 31:97--105, 1983.

\bibitem{fritschel:ISC}
P.~Fritschel, R.~Bork, G.~Gonzalez, N.~Mavalvala, D.~Ouimette, H.~Rong,
  D.~Sigg, and M.~Zucker.
\newblock {\em Applied Optics}, 40:4988--4998, 2001.

\bibitem{mevans:locking}
M.~Evans, N.~Mavalvala, P.~Fritschel, R.~Bork, B.~Bhawal, R.~Gustafson,
  W.~Kells, M.~Landry, D.~Sigg, R.~Weiss, S.~Whitcomb, and H.~Yamamoto.
\newblock {\em Optics Lett.}, 27:598--600, 2002.

\bibitem{daw:MSFF}
J.~A. Giaime, E.~J. Daw, M.~Weitz, R.~Adhikari, P.~Fritschel, R.~Abbott,
  R.~Bork, and J.~Heefner.
\newblock {\em Rev. Sci. Instrum.}, 74, 2003.

\bibitem{daw:SE}
E.~J. Daw, J.~A. Giaime, D.~Lormand, M.~Lubinski, and J.~Zweizig.
\newblock {\em Class. Quantum Grav.}, 21:2255, 2004.

\bibitem{sjw:scatter}
D.~J. Ottaway, P.~Fritschel, and S.~J. Waldman.
\newblock {\em Optics Express}, 20:8329--8336, 2012.

\bibitem{wen:HEPI}
S.~Wen.
\newblock PhD thesis, Louisiana State University, 2009.

\bibitem{hua:FIR}
W.~Hua, D.~B. Debra, C.~T. Hardham, B.~T. Lantz, and J.~A. Giaime.
\newblock {\em Proceedings of ASPE Spring 2004 Topical Meeting on Control of
  Precision Systems}, 2004.

\bibitem{rosenband:FPphasenoise}
M.~J. Thorpe, D.~R. Leibrandt, T.~M. Fortier, and T.~Rosenband.
\newblock {\em Opt. Express}, 18:18744, 2010.

\bibitem{jenne:40m}
J.~C. Driggers, M.~Evans, K.~Pepper, and R.~Adhikari.
\newblock {\em Rev. Sci. Instrum.}, 83, 2012.

\bibitem{norbert}
N.~Wiener.
\newblock {\em {Extrapolation, Interpolation, and Smoothing of Stationary Time
  Series}}.
\newblock M.I.T. Press, 1964.

\bibitem{mathworks}
MathWorks.
\newblock {MATLAB 2010}.

\bibitem{durbin}
J.~Durbin.
\newblock {\em Rev. Inst. Int. Stat.}, 28, 1960.

\bibitem{huang}
Y.~Huang, J.~Benesty, and J.~Chen.
\newblock {\em Acoustic {MIMO} {S}ignal {P}rocessing}.
\newblock Signals and Communication Technology. Springer, 2006.

\bibitem{VectfitPaper}
B.~Gustavsen and A.~Semlyen.
\newblock {\em IEEE Transactions on Power Delivery}, 14:1052--1061, 1999.

\bibitem{aLIGO:SEI}
R.~Abbott, R.~Adhikari, G.~Allen, S.~Cowley, E.~Daw, D.~DeBra, J.~Giaime,
  G.~Hammond, M.~Hammond, C.~Hardham, J.~How, W.~Hua, W.~Johnson, B.~Lantz,
  K.~Mason, R.~Mittleman, J.~Nichol, S.~Richman, J.~Rollins, D.~Shoemaker,
  G.~Stapfer, and R.~Stebbins.
\newblock {\em Classical and Quantum Gravity}, 21(5):S915, 2004.

\bibitem{hardham:actuators}
C.~Hardham.
\newblock {\em Proceedings of ASPE Spring 2004 Topical Meeting on Control of
  Precision Systems}, 2004.

\bibitem{andri:PRCoffset}
A.~M. Gretarsson, E.~D'Ambrosio, V.~Frolov, B.~O'Reilly, and P.~K. Fritschel.
\newblock {\em J. Opt. Soc. Am. B}, 24, 2007.

\bibitem{chatterji:Q}
S.~Chatterji.
\newblock PhD thesis, Massachusetts Institute of Technology, 2005.

\bibitem{blantz:rotations}
B.~Lantz, R.~Schofield, B.~O'Reilly, D.~E. Clark, and D.~DeBra.
\newblock {\em Bulletin of the Seismological Society of America}, 2009.

\end{thebibliography}

\end{document}